# Selective Hybridization of a Terpyridine-Based Molecule with a Noble Metal

*M. Capsoni[||, ‡], A. Schiffrin[||, ‡,†], K. A. Cochrane[#], C.-G. Wang[^], T. Roussy[||], A. Q. Shaw[||], W. Ji[^], S. A. Burke[||,#,~,*]*

|| Department of Physics and Astronomy, University of British Columbia, Vancouver British Columbia, Canada, V6T 1Z1

# Department of Chemistry, University of British Columbia, Vancouver British Columbia, Canada, V6T 1Z1

^ Department of Physics and Beijing Key Laboratory of Optoelectronic Functional Materials & Micro-nano Devices, Renmin University of China, Beijing 100872, People's Republic of China

~ Quantum Matter Institute, University of British Columbia, Vancouver British Columbia, Canada, V6T 1Z4

‡ these authors contributed equally

* corresponding author: saburke@phas.ubc.ca

† current address: School of Physics & Astronomy, Monash University, Clayton, Victoria 3800, Australia

ABSTRACT:

The electronic properties of metal-molecule interfaces can in principle be controlled by molecular design and self-assembly, yielding great potential for future nano- and opto-electronic technologies. However, the coupling between molecular orbitals and the electronic states of the surface can significantly influence molecular states. In particular, molecules designed to create metal-organic self-assembled networks have functional groups that by necessity are designed to interact strongly with metals. Here, we investigate the adsorption interactions of a terpyridine (tpy)-based molecule on a noble metal, Ag(111), by low-temperature scanning tunneling microscopy (STM) and spectroscopy (STS) together with density functional theory (DFT) calculations. By comparing the local density of states (DOS) information gained from STS for the molecule on the bare Ag(111) surface, with that of the molecule decoupled from the underlying metal by a NaCl bilayer, we find that tpy-localized orbitals hybridize strongly with the metal substrate. Meanwhile, those related to the phenyl rings that link the two terminal tpy groups are less influenced by the interaction with the surface. The selective hybridization of the tpy groups provides an example of strong, orbital-specific electronic coupling between a functional group and a noble metal surface, which may alter the intended balance of interactions and resulting electronic behavior of the molecule-metal interface.

INTRODUCTION:

Controlling the electronic properties of organic molecules, especially those with extensive π-conjugation, adsorbed on metals holds promise for the development of molecular-based electronic[1-3], spintronic[4-9] and optoelectronic[6,9,10] nanotechnologies. Many such applications take advantage of the ability to tailor the electronic properties of molecules by tuning their size, symmetry, shape and attached functional groups.[7,11] One method that allows a high degree of control and tuning of the metal-molecule interface structure is on-surface self-assembly, where the interactions between molecular building blocks, or "tectons", are designed to drive the formation of particular motifs over large areas.[4-9] The directional interactions required for such self-assembly processes typically rely on specific functional groups that provide non-covalent interactions between molecules, such as hydrogen-bonding, or between the organic components and metal atoms, where coordination bonds are typically used to link the network together. For this latter case of metal-organic self-assembly[6,9,10], the functional groups must be selected specifically to have a strong interaction with metals. As the self-assembly process relies on a fine balance of molecule-molecule and molecule-substrate interactions[7,11], understanding how these chelating functional groups interact with the underlying metal surface is also an essential component of the molecular design. Further, in the design of functional molecule-metal interfaces the influence of the metal surface may outweigh the influence of functional groups designed to tune molecular states. To obtain metal-organic networks with reliably predictable electronic properties, one must understand the whole molecule-metal interface.

Here, we study an organic molecule with terpyridine functional groups that can serve as a basis for metal-organic self-assembly on surfaces. Polypyridyl ligands are frequently used as ligands in metal-organic complexes[12], coordination polymers[13,14], and have recently been used as tectons for self-assembly[15]. These systems exhibit a wide range of interesting photophysical, electrochemical, and magnetic properties when bound to transition metals[12], and a large number of different compounds have been designed for various applications such as membranes[16], gas storage[17], photoluminescence[18], solar energy conversion[19,20], catalysis[21-23], nanoelectronics[24], and spin-crossover systems[25]. In this work, we use scanning tunneling microscopy (STM) and spectroscopy (STS), together with density functional theory (DFT) calculations, to investigate the adsorption geometry and electronic states of terpyridine-phenyl-phenyl-terpyridine (TPPT) molecules (Scheme 1) deposited on Ag(111). We compare the spatially and energy resolved electronic state information from STS of the molecules on both the bare Ag surface and decoupled from the metal by an NaCl bilayer[26,27]. This comparison between weak and strong coupling to a metal allows us to obtain information on the orbital specific interaction with the metal surface.

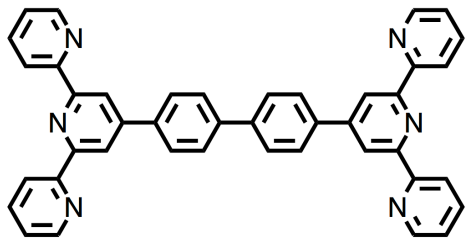

Scheme 1: Terpyridine-phenyl-phenyl-terpyridine (TPPT) chemical structure.

METHODS:

**Sample preparation.** Ag(111) surfaces (Mateck GmbH) were prepared in ultra high vacuum (UHV) using repeated cycles of Ar$^+$ sputtering and annealing at 790 K. To electronically decouple the molecules from the silver substrate[26,27], NaCl (TraceSELECT ≥99.999%, Fluka) was deposited at ~800K on the clean Ag(111) held at ~340K, resulting in (001) bilayer islands covering ~ 40% of the surface. TPPT molecules (HetCat Switzerland; see Scheme 1) were sublimed at 550 K onto the bare Ag(111) substrate held at room temperature (RT), or onto the NaCl/Ag(111) surface held at ~4.3K. Typical depositions were performed at rates of ~ 4 x 10$^{-4}$ molecules/(nm$^2$ s) at pressures of < 2x10$^{-9}$ mbar during evaporation.

**Scanning tunneling microscopy and spectroscopy.** STM and STS measurements were performed in UHV (1 x 10$^{-10}$ mbar) at ~4.3K (Omicron Nanotechnology) with a Ag-terminated Pt/Ir tip. Reference spectra on the bare Ag(111) surface were used to verify that the tip was metallic. STM imaging was performed in constant-current mode. The sample bias is reported throughout the text. Spectroscopy measurements were performed by positioning the tip on a fixed point on the sample, disabling the feedback loop, and measuring the tunneling current ($I_t$) at a function of bias voltage ($V_b$) with an energy resolution of at least 10meV. STS measurements were acquired either at specific positions, or as grids of I(x,y,V) with a spatial resolution of 1.9pm/pixel. (dI/dV)/(I/V) spectra were obtained by taking the numerical derivative and dividing by (I/V) to reduce the exponential background arising from the transmission function at larger tip-sample biases.[28] STS curves from the grids were averaged over the areas indicated in the figures, and STS maps were produced from the (x,y) grid data at the indicated energy.

**Density functional theory details.** Structural calculations of TPPT molecules on the Ag(111) surface were carried out using Density functional theory (DFT) via the Projector Augmented Waves (PAW) method[29,30], and a plane-wave basis set as implemented in the Vienna Ab-initio Simulation Package (VASP)[31,32] with the General Gradient Approximation (GGA) taken for the exchange-correlation potentials. The revised Perdew-Burke-Ernzerhof (RPBE)[33] functional, with the second version of Grimme's dispersion corrections (DFT-D2)[34], was adopted. Treatment of van der Waals interactions at the DFT-D2 level gives a better description of geometries and corresponding energies for molecule-metal systems than those from the standard DFT[35]. The kinetic energy cutoff for the plane wave basis was set to 400 eV in configuration optimizations. An (11 x 7) supercell consisting of 308 Ag atoms in 4-layers was used. Only the gamma point of the surface Brillouin zones were sampled for geometry optimizations. All atoms except those at the bottom two Ag layers were fully relaxed until the residual force per atom was less than 0.02 eV/Å.

The molecular orbitals and energy levels of an isolated TPPT, i.e. not including a surface, were computed using the Gaussian 09 computer package on a fully relaxed TPPT molecule in the gas-phase (i.e. neglecting interactions with other molecules as well as with a surface) using the B3LYP/6-31g basis set[36]. The isosurface level used for electronic density plots was 0.02 electrons/$r_{Bohr}^3$. It has been previously demonstrated for molecules adsorbed on NaCl, that calculations of the electron density of the free molecule (neglecting surface interactions) qualitatively agree well with the observed STS contrast of the molecular orbitals[26,27]

RESULTS AND DISCUSSION:
In STM images the TPPT molecules deposited on the Ag(111) surface appear as "dog-bone" shaped features [Figure 1 and Fig. 2(a)]. At low coverages (significantly less than one monolayer) TPPT is observed as either isolated molecules or adjacently bonded following staggered rows or zigzag patterns [Figure 1]. Careful examination of the relative alignment of the molecules show six specific orientations indicated by the red and yellow labels in Fig. 1(a). At higher coverages, close-packed tiled domains of the staggered rows emerge [Fig. 1(b)].

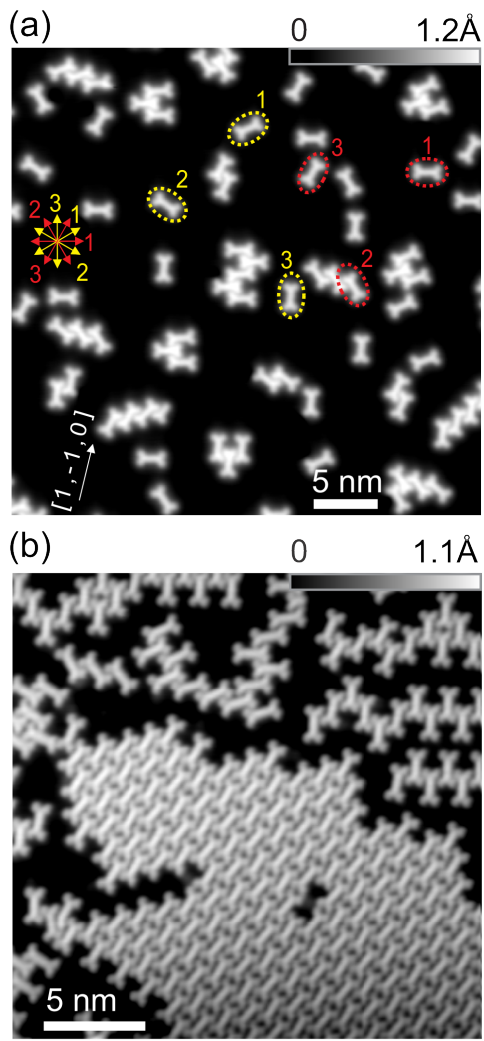

Figure 1: Constant-current STM images of (a) low coverage ($V_b$ = 200mV, $I_t$= 50pA) and (b) high coverage ($V_b$ = -500mV, $I_t$= 20pA) depositions of TPPT molecules on Ag(111) (substrate at room temperature). Molecules adsorb following ~ ± 16° angle with respect to the <1, 1, 0> crystallographic directions (white arrow), resulting in six equivalent orientations denoted by labels 1, 2 and 3, red (CW) and yellow (CCW) dashed ellipses and arrows. At high coverage, close-packed regions emerge with a head-to-head arrangement of the tpy groups that is not seen at low coverage.

Atomic resolution images of the Ag(111) surface alongside isolated TPPT molecules show that the molecule adsorbs at an angle of ±(16±1)° relative to the <1,1,0> directions of the Ag surface [Fig. 2(a)]. This experimental result is also supported by DFT calculations of the minimum energy configuration of the molecule-metal system [Fig. 2(b)]. This adsorption geometry results in 6 possible orientations corresponding to the two mirror symmetric reflections about the Ag(111) lattice directions, all of which were observed in large scale images [see e.g. Fig. 1(a)] confirming the adsorption configuration determined by high resolution imaging. The TPPT molecules also appear flat in the STM images, similar to other terpyridine (tpy) containing molecules examined by STM[37-39] and DFT[40]. Our DFT calculations indicate a small twist angle between the two phenyls [Fig. 2(c)] in the middle of the molecule breaking the planarity, which is sometimes subtly observed in the STM images. The well-defined orientations of the molecules and their near-planar adsorption indicate a significant molecule-substrate interaction, as expected for π-conjugated molecules on metals, particularly for N-heterocycles such as pyridines[41].

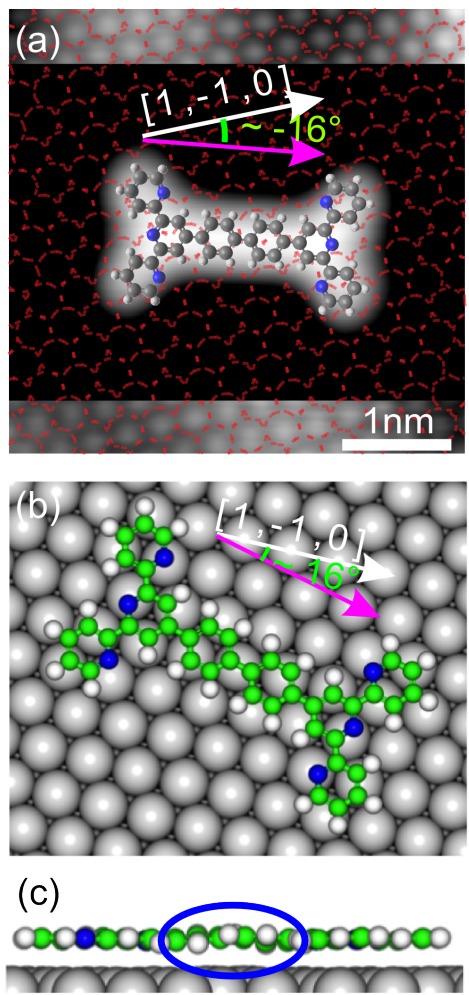

Figure 2: (a) Constant-current STM image resolving both the isolated molecule and the underlying Ag(111) substrate. The feedback parameters were changed during the scan to allow for the close tip-

sample distances needed for atomic resolution of the metal and the larger tip-sample distances required to not perturb the molecule. (parameters for molecule: $V_b$ = -200mV, $I_t$ = 10pA; for Ag(111): $V_b$= -10mV, $I_t$ = 300nA). Chemical structure of the molecule and the silver lattice are superimposed on the STM image. The orientation of the molecule (pink vector) with respect to the [1, -1, 0] direction (white vector) is shown. (b) Minimum energy configuration of TPPT adsorbed on Ag(111) calculated by DFT-D2, with the side-view (c) showing the small distortion of the two central phenyl rings.

High-resolution STM images of TPPT molecules at small negative biases show slight depressions next to the outer pyridine rings of the tpy groups [red arrows in Figs. 3(a) and (b)]. These depressions are indicative of a reduction of the silver local density of states (LDOS) likely caused by an electrostatic repulsion of the conduction electrons in the silver surface by the electronegative N atoms of the pyridine. This indicates that the N atoms of the outer pyridine rings point away from the tpy group in both the isolated molecule and the dimer, consistent with the gas-phase configuration (as in Scheme 1). This configuration also lends insight into the adjacent bonding that gives rise to the staggered rows and zig-zag patterns observed in Figure 1: the outward-oriented N of the peripheral pyridine groups can interact via an attractive proton acceptor/organic ring interaction[42,43] with the hydrogen of the adjacent molecule [green dashed ovals in Fig. 3(a)]. Hence, the tpy groups remain in their gas phase configuration when adsorbed on the surface and mediate the intermolecular interactions that drive the observed packing on Ag(111).

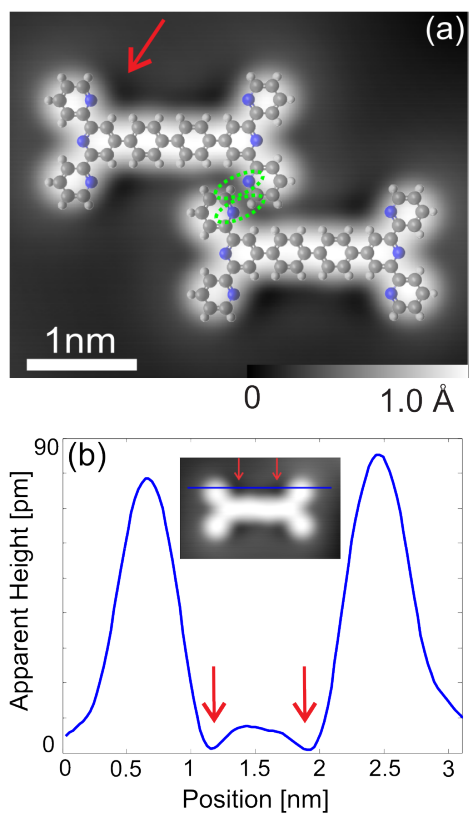

Figure 3: (a) Constant-current STM image ($V_b = -10mV$, $I_t = 1nA$) of two TPPT molecules on Ag(111) adjacently bound by attractive proton acceptor/organic ring interactions (dashed green ellipses). (b) Apparent height profile, indicated in topography inset, of an isolated TPPT on Ag(111) ($V_b = 10mV$, $I_t = 1nA$). Red arrows indicate the depressions observed next to the outer pyridine groups for both the dimer and the isolated TPPT.

To investigate the electronic states that may be involved in the adsorption interaction with the Ag(111) substrate, STS measurements were performed on TPPT molecules on the bare Ag(111) surface and a bilayer of NaCl on Ag(111). This thin insulating layer reduces the electronic interaction of adsorbed molecules with the underlying metal allowing STM/STS measurements of molecular states that are close to those expected for the free, isolated molecule[26,27]. Note that for TPPT molecules on NaCl, very low tunneling current set points (typically a few pA) were needed to avoid perturbing the molecules leading to lower signal-to-noise both in topographic images and STS measurements. This also resulted in a limited range of accessible biases compared with molecules on Ag(111).

Figure 4a and b show STS measurements over the middle and tpy ends of the molecule. Within the experimentally accessible energy range several features can be noted: (i) there is no obvious occupied state ($V_b<0$) signature observed within the accessible range of biases on either substrate; (ii) on Ag(111) there are two prominent tunneling resonances over the center of the molecule, at $V_b$=+1.15V and +2.3V, while only weak features are observed over the tpy ends; and (iii) on the NaCl bilayer on Ag(111) there are three prominent tunneling resonances, at $V_b$ =+1.65V, +2.2V and >3V, over the center of the molecule with the first two replicated at the tpy end with similar intensity (the third peak was inaccessible). Comparing the energy spacings of the states observed on the NaCl bilayer to the weak features seen on the tpy ends directly on Ag(111), there is a slight bump between the two more prominent peaks that may correspond to the same state. The differences between the STS results on Ag(111) compared to those on NaCl, which is expected to be weakly interacting, indicate a significant electronic interaction of TPPT with the metal surface. To explore the evolution of these states spatially and in energy, both simulations and more detailed measurements of the spatial distribution were performed.

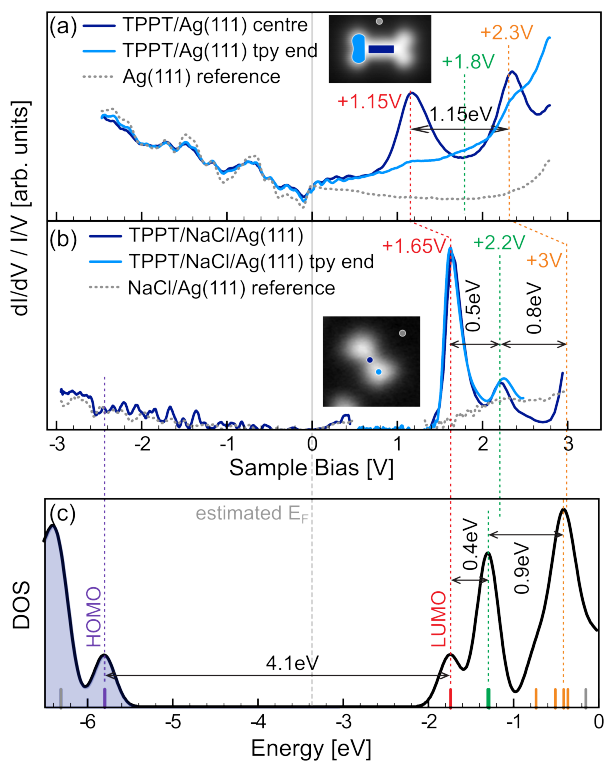

Figure 4: (a) (dI/dV)/(I/V) spectra of an isolated TPPT/Ag(111): dark blue averaged over middle region, light blue averaged over tpy end, and grey reference of Ag(111) (see inset for STS positions). Grid and topography set-point parameters: $V_b$ = -2.50V, $I_t$ = 50pA. (b) (dI/dV)/(I/V) spectra of an isolated molecule

on NaCl/Ag(111) obtained from multiple STS measurements taken in the same position with different set-point parameters and bias ranges; scaled for comparison: dark blue taken at centre of molecule, light blue taken at tpy end, and grey on the NaCl/Ag(111) for reference (see inset for STS positions). (c) Calculated density of states (DOS) for an isolated, gas-phase TPPT, with energy referenced to the vacuum level; the spectra are plotted to align the first tunneling resonance on NaCl with the first state in the DFT, grey vertical line indicates approximate Fermi energy based on this alignment. Vertical bars and lines indicate energies of states discussed: purple, HOMO; red, LUMO; green, LUMO+1/+2/+3; orange, cluster of states above LUMO +4, grey bars show other states not discussed.

As the NaCl bilayer serves to electronically decouple the molecule from the metal substrate, DFT simulations for an isolated, gas-phase TPPT molecule (not including a substrate) were performed to qualitatively identify the orbitals based on shape, symmetry, nodes, and relative energy positions and compare with experimental STS [Figs. 4(a), 4(b)]. Fig. 4(c) shows the DFT-calculated density of states (referenced to the vacuum level), with the Highest Occupied Molecular Orbital (HOMO) located at -5.85eV (marked with the purple line), and Lowest Unoccupied Molecular Orbital (LUMO) at -1.73eV (marked with the red line), followed by two other prominent peaks at higher energies corresponding to the LUMO+1/+2/+3 (green vertical line) and a cluster of states with a first shoulder at the LUMO+4 (orange vertical line). By aligning the LUMO from the DFT calculation to first resonance observed at positive bias (unoccupied states) in the STS of TPPT/NaCl (2 ML)/Ag(111), an occupied state tunneling resonance was expected around -2.5V corresponding to the HOMO [see Fig. 4(b),(c)]. Although the HOMO was not observed in STS measurements, the B3LYP-6-31g functional used in the calculation often underestimates the HOMO-LUMO gap[44], likely placing the HOMO deeper than was accessible to the STS measurement. However, at positive bias, the three peaks in the STS for TPPT/NaCl (2 ML)/Ag(111) correspond quite well with the unoccupied DOS calculated by DFT, both in terms of the number of peaks and the energy differences between them. Now examining the STS for TPPT directly on Ag(111) [Fig. 5(a)], in addition to a downward shift in energy of the states observed, the energy difference between the two peaks observed correspond more closely to the 1$^{st}$ and 3$^{rd}$ peaks observed on bilayer NaCl. To more clearly visualize the progression of each of these unoccupied states, STS taken across the length of the molecule only at positive bias are shown in Fig. 5(a) and (b) for the NaCl bilayer and bare Ag(111) surfaces respectively.  Here, while the first peak in the STS on NaCl is strongest in the middle, one can see that the 2nd STS peak is more prominent on the tpy ends and slightly fades towards the center of the molecule.  On the bare Ag(111), there is a slight shoulder that only appears at the tpy ends of the molecule but is never pronounced, while the other states are prominent in the middle, and the third peak also has intensity over the tpy groups.  Notably, the same molecule was previously studied on the Cu(111) surface showing similar spectra for the center position[39], but with a distinguishable peak on the tpy end between the peaks that were also seen at the middle of the molecule.  Here, there is a slight increase in the STS intensity over the tpy end compared to the middle at around +1.8V, but no prominent peak is observed.  This difference compared to the Cu surface, which the authors found to be weakly interacting via DFT calculations, and our comparison with spectra on the molecule adsorbed on NaCl point toward a "missing" state in the STS measurements over the tpy groups indicating broadening of this state.  The significant broadening of this peak over the tpy ends where the LUMO+1/+2/+3 orbitals are most strongly localized indicates hybridization with the surface via a strong interaction of the pyridine rings with the Ag(111) substrate.

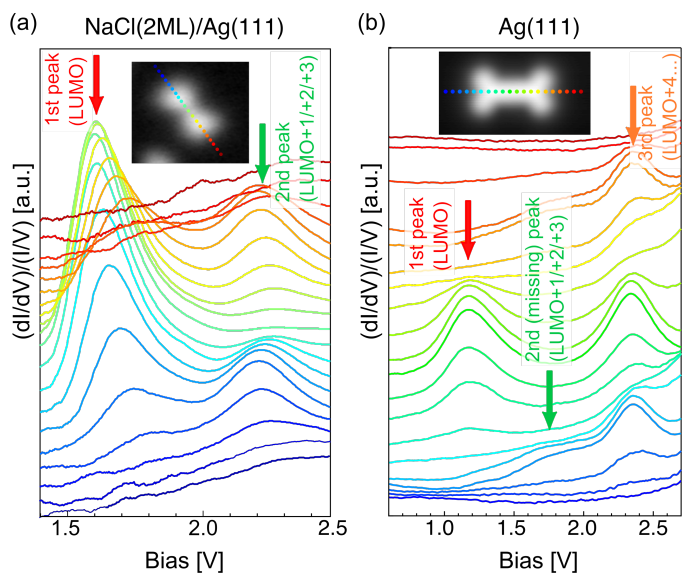

**Figure 5**: STS across isolated TPPT molecule on (a) NaCl/Ag(111), and (b) directly on Ag(111) showing the progression of each peak across the molecule.  Peaks are labelled with arrows: 1st peak (LUMO) = red, 2nd peak (LUMO+1/+2/+3) = green, 3rd peak (LUMO+4 to +7) = orange.  Notably the 2nd peak associated with the LUMO+1/+2/+3 is more prominent over the tpy ends of the molecule, and is barely discernible on the TPPT directly on Ag(111).

In order to more clearly identify the states corresponding to the peaks in STS on TPPT for each surface (Ag and bilayer salt), we plotted the (dI/dV)/(I/V) intensity along along the long molecular axis through the middle of the molecule at the energies discussed above (see Fig. 6). The (dI/dV)/(I/V) intensity profiles for a TPPT on silver were obtained from STS grid measurements and the full map is also shown in the inset. For isolated TPPT on salt, grid measurements over large bias ranges were not obtainable due to frequent instability of the molecule disrupting these longer measurements, hence the profiles were instead obtained from a single line of point spectra except for the first unoccupied peak.

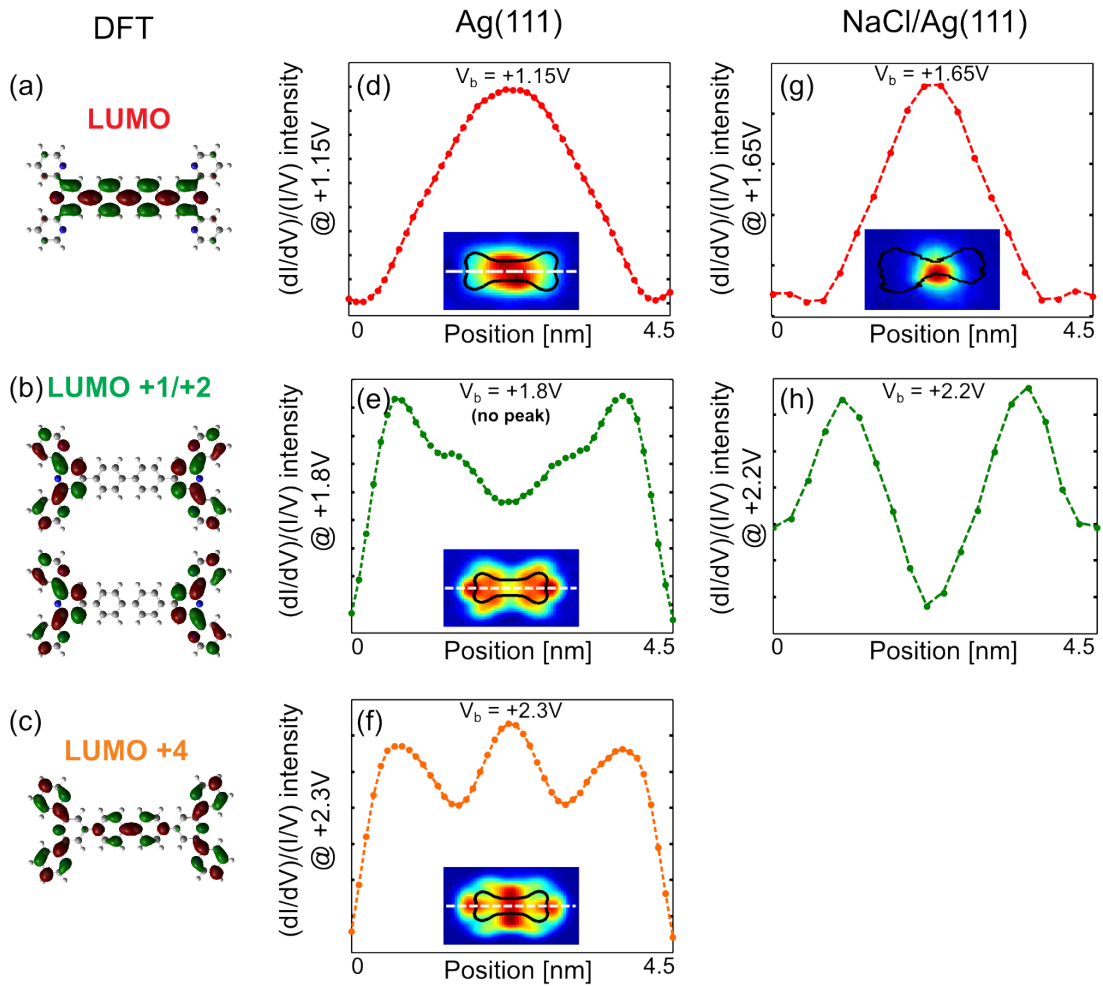

Figure 6: (a) – (c) Electron density maps calculated by DFT for an isolated TPPT molecule (isosurface 0.02 electrons/$r_{Bohr}^3$) of the (a) LUMO, (b) the degenerate LUMO+1/+2, and (c) LUMO+4. (d) – (f) STS profiles along the long axis of an isolated TPPT on Ag(111) (white dashed lines in insets) at (d) +1.15V, (e) +1.8V, and (f) +2.3V. Insets: STS maps at corresponding bias voltages (grid set-point parameters:

$V_b$=-2.5V, $I_t$ = 50pA) with overlaid molecular contour (black outline). (g) and (h) STS profiles of an isolated TPPT on NaCl/Ag(111) at (g) +1.65V and (h) +2.2V. Inset in (g): STS map of an isolated TPPT/NaCl (2 ML)/Ag(111) (set-point parameters: $V_b$=-2.0V, $I_t$ = 0.6pA). A full spatial map was not possible for TPPT/NaCl (2 ML)/Ag(111) at 2.2V or above.

Comparing the profiles at the energies of the first (unoccupied) peaks for TPPT molecules on each substrate, the STS intensity profile at +1.15V of TPPT/Ag [Fig. 6(d)] matches that at +1.65V of a TPPT/NaCl (2 ML)/Ag [Fig. 6(g)] quite well. Both show the highest intensity at the middle of the long axis, as is also clearly shown in the STS maps at these biases [insets of the two panels (d) and (g)]. This is also consistent with the electronic density calculated for the LUMO, which is centered along the axis of the molecule without significant density on the tpy groups. However, the +2.2V STS intensity profile of a TPPT on NaCl (2 ML)/Ag [Fig. 6(h)] is clearly distinct from the +2.3V profile (at the next peak) for TPPT on Ag [Fig. 6(f)]. Fig. 6(h) shows two symmetric humps about the middle of the molecule, whereas Fig. 6(f) shows three less pronounced humps: two at the ends (on the tpy groups) and one in the middle of the TPPT molecule. These distinct profiles indicate that they do not represent the same spatial electronic distribution, and therefore these two tunneling resonances do not correspond to the same MO. Additional evidence supporting this statement is found in the energy difference ($\Delta V$) between the first two peaks: on Ag the $\Delta V$ = 1.15V, whereas on bilayer NaCl the $\Delta V$ = 0.5V which is similar to the $\Delta V$=0.4V from the DFT calculation. Alternatively, the +2.3V peak on Ag could correspond to the +3V peak on NaCl, giving more similar energy differences relative to the first peak with $\Delta V$=1.15V (Ag), and $\Delta V$=1.35V (NaCl). Acquisition of a (dI/dV)/(I/V) line profile of TPPT/NaCl (2 ML)/Ag at ≥+3V to verify this experimentally was not possible due to instability of the TPPT molecules on the bilayer salt at this bias. However, the DFT-calculated electronic distribution of the LUMO+4 [Fig. 6(c)] seems to support this supposition: the isosurface map of electron density shows that this state is localized both on the tpy groups and around the center of the molecule with nodes between. This, combined with other MOs in a similar energy range (note there are several states contributing to the peak around this energy) could reasonably produce the intensity profile observed by STS in Fig. 6(f). Thus, we attribute the first peak at positive bias on both TPPT/Ag and TPPT/NaCl (2 ML)/Ag to the LUMO, and the highest energy peak observed to the LUMO+4 ($V_b$=2.3V for Ag, and $V_b$>3V for NaCl/Ag)

To search for evidence of the LUMO+1/+2 for TPPT on Ag, we use the energy differences found for TPPT/NaCl (2 ML)/Ag and the DFT calculation as a guide and examine the (dI/dV)/(I/V) intensity profiles between the two clear tunneling resonances observed. As noted above, there is a slight increase in intensity on the tpy ends compared to the middle of the molecule in this energy range, and at $V_b$=+1.8V we find an intensity profile of TPPT/Ag [Fig. 6(e)] that reasonably matches the profile of the +2.2V peak from TPPT/NaCl (2 ML)/Ag [Fig. 6(h)]. In this case, the energy difference from the first resonance to where this profile is found is $\Delta V$=0.65V, which is reasonably close to the $\Delta V$=0.5V found on NaCl from the 1$^{st}$ to the 2$^{nd}$ peak. These energy differences are also consistent with the LUMO to LUMO+1/+2 calculated from DFT. The LUMO+1/+2 state corresponds to an electron density mostly localized on the tpy groups [Fig. 6(g)], giving rise to the profiles showing two bumps towards the ends of the molecule. Note: the LUMO+3 is also close in energy to the LUMO+1/+2 and also mostly found on the tpy groups with some additional density on the phenyl groups. As tpy groups are expected to interact strongly with metals[37,40,41] this part of the molecule is the most likely to show strong hybridization with the silver substrate. The lack of a pronounced peak in STS at the position of this MO indicates significant broadening of the tpy-associated orbital due to this hybridization.

CONCLUSION:

Here, we have used STM and STS, supported by DFT calculations, to study the interaction of a bis-terpyridine molecule with a metal surface, Ag(111). Both the adsorption configuration and electronic structure of the TPPT molecules appear to be strongly influenced by the interaction with the silver substrate. STS measurements revealed that the electronic structure of this tpy-based molecule is substantially altered upon adsorption on the Ag(111) surface, by comparing to the same molecule on a weakly interacting surface (NaCl). In particular, the number of prominent tunneling resonances associated with MOs and their energies differed significantly on the two substrates. A molecular state strongly localized on the tpy groups identified by spatial mapping is substantially broadened, making it nearly indistinguishable from the background in STS, indicating strong hybridization between this particular state and the substrate electrons. Previous work on the same, and similar molecules on Cu(111)[39] indicate a weaker interaction with the surface further emphasizing the need to understand the interaction of key functional groups with various metal substrates.

These results indicate that specific functional groups may interact strongly and hybridize with metal surfaces resulting in highly delocalized metal-molecule states while leaving other states relatively distinct, as has been seen for example in cases where interface states between adsorbed molecules and noble metal surfaces are formed[45]. This is a particular concern for tectons targeted towards metal-organic self-assembly on metallic surfaces since by design the functional groups must strongly interact with metal atoms. Here, this was found to be the case for tpy groups, part of a key class of compounds for organic photovoltaic and catalysis applications; applications where interfaces between the organic and a solid support such as a metal or semiconductor are needed or desired. Further, the coupling between specific electronic states localized on a particular functional group of a larger molecule could also be exploited as a new method of optimizing molecule-metal interfaces with sub-molecular precision.

ACKNOWLEDGMENTS: The authors gratefully acknowledge support from the NSERC Discovery grants program No. 402072-2012, ACS PRF 55955-ND, the Canadian Foundation for Innovation, the NSERC CREATE-QUEST grant No. 414110-12 (T.R. and S.B.), the Canada Research Chairs Program (S.B.), the University of British Columbia 4-year Fellowship (M.C.), A. S. was supported by the Max Planck-UBC Centre for Quantum Materials and acknowledges the ARC Centre of Excellence for Future Low-Energy Electronics Technologies for ongoing support. This work was also supported by the National Natural Science Foundation of China (NSFC) under Grant Nos. 11274380, 91433103, 11622437 and 61674171, the Ministry of Science and Technology (MOST) of China under Grant No. 2012CB932704, and C.-G. W. was supported by the Outstanding Innovative Talents Cultivation Funded Programs 2015 of Renmin University of China. The geometry optimization Calculations were performed at the Physics Laboratory for High-Performance Computing of Renmin University of China and at the Shanghai Supercomputer Center.

ASSOCIATED CONTENT

Supporting information is available online:

Electron density plots of unoccupied molecular orbitals contributing to the notable peaks in the total DOS calculated with Gaussian

TOC graphic

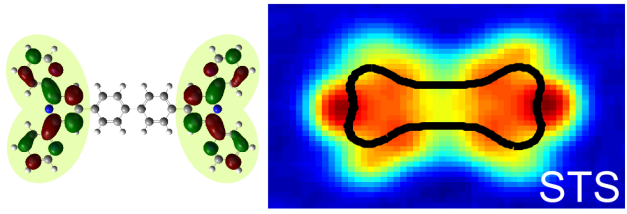